\documentclass[]{aa}       % New LaTeX A&A Standard Fonts
\usepackage{graphicx}
\usepackage{amssymb}
\usepackage{longtable}
\usepackage{supertabular}
\usepackage{natbib} 
\usepackage{ulem}

\begin{document}

\title{Non-thermal emission from standing relativistic shocks:\\ an application to red giant winds interacting with AGN jets}

\author{
V. Bosch-Ramon \inst{1}
}

\authorrunning{Bosch-Ramon, V.}

\titlerunning{Non-thermal emission from standing relativistic shocks}

\institute{
Departament d'Astronomia i Meteorologia, Institut de Ci\`encies del Cosmos (ICC), Universitat de Barcelona (IEEC-UB), Mart\'i i
Franqu\`es 1, 08028 Barcelona, Catalonia, Spain
}

\offprints{V. Bosch-Ramon, \email{vbosch@am.ub.es}}

\date{Received <date> / Accepted <date>}

\abstract
{Galactic and extragalactic relativistic jets are surrounded by rich environments that are full of moving objects, such as stars and dense medium inhomogeneities. These objects can enter into the jets and generate shocks and non-thermal emission.} 
{We characterize the emitting properties of the downstream region of a standing shock formed due to the interaction of a relativistic jet with an obstacle. We focus on the case of red giants interacting with an extragalactic jet.}  
{We perform relativistic axisymmetric hydrodynamical simulations of a relativistic jet meeting an obstacle of very large inertia. The results are interpreted in the framework of a red giant whose dense and slow wind interacts with the jet of an active galactic nucleus. Assuming that particles are accelerated in the standing shock generated in the jet as it impacts the red giant wind, we compute the non-thermal particle distribution, the Doppler boosting enhancement, and the non-thermal luminosity in gamma rays.}    
{The available non-thermal energy from jet-obstacle interactions is potentially enhanced by a factor of $\sim 100$ when accounting for the whole surface of the shock induced by the obstacle, instead of just the obstacle section. The observer gamma-ray luminosity, including the effective obstacle size, the flow velocity and Doppler boosting effects, can be $\sim 300\,(\gamma_{\rm j}/10)^2$ times higher than when the emitting flow is assumed at rest and only the obstacle section is considered, where $\gamma_{\rm j}$ is the jet Lorentz factor. For a whole population of red giants inside the jet of an active galactic nucleus, the predicted persistent gamma-ray luminosities may be potentially detectable for a jet pointing approximately to the observer.}{Obstacles interacting with relativistic outflows, for instance clouds and populations of stars for extragalactic jets, or stellar wind inhomogeneities in microquasar jets and in winds of pulsars in binaries, should be taken into account when investigating the origin of the non-thermal emission from these sources.} 

\maketitle

\section{Introduction}\label{intro}

Relativistic jets are surrounded by rich environments, and stars and dense medium inhomogeneities can enter into these jets leading to the formation of shocks, as considered for instance in \cite{blandford1979,coleman1985,komissarov1994,bednarek1997,araudo2009,araudo2010,araudo2013,barkov2010,barkovAAA,perucho2012}; and \cite{bosch-ramon2012}. These shocks are, only initially in some cases, almost standing shocks, at rest in the laboratory frame (LF). A similar situation can occur in the termination region of other relativistic outflows, such as pulsar winds in binary systems \citep{paredes-fortuny2015}. 

In several previous papers that modelled the radiation from relativistic shocks at rest in the LF, the emitting region was circumscribed to the closest obstacle vicinity and, as the flow there is only mildly relativistic, the radiation Doppler boosting was not considered. However, there could be more jet energy available for non-thermal particles than assumed in these works, as the effective interaction area may be much larger than just the obstacle section. In addition to that, the flow initially moves at $\approx c/3$ right after the shock, and then is quickly accelerated further downstream because of strong pressure gradients. This effect is seen for instance in fig. 2 in \cite{bosch-ramon2012}. Therefore, even relatively close to the obstacle at rest, the shocked flow velocity is high enough for Doppler boosting to significantly enhance radiation in the emitting fluid direction of motion. The motion of the shocked flow could be also considered for an obstacle that is itself accelerated by the jet ram pressure \citep[e.g.][]{barkovBBB,khangulyan2013}, but this would require accounting for the complex dynamical evolution of the obstacle \citep[see][for hydrodynamical simulations predicting the obstacle disruption]{bosch-ramon2012}. 

In this work, we study the kinetic-to-internal (possibly non-thermal) energy conversion, gamma-ray radiation efficiency, and Doppler boosting effects, in the shock and the shocked flow of a relativistic hydrodynamic jet interacting with an obstacle at rest. The obstacle could be a star or a cloud in the case of an extragalactic jet, or a stellar wind inhomogeneity in the case of a galactic microquasar jet.  Since this is a relevant and illustrative case, we focus here on the case of a population of red giant stars (RG) interacting with the jet of an active galactic nucleus (AGN). We first characterize the non-thermal luminosity from the shocked jet region accounting for hydrodynamical information, and second we adopt a phenomenological description of the spatial distribution of the RG in the regions crossed by the jet as it goes through the galaxy.  To compute the emission, we consider mainly inverse Compton (IC) scattering, assume that synchrotron radiation is a minor emission channel, and put special emphasis on the GeV-TeV range. Studies such as this are needed to characterize the non-thermal emission from interactions between stars and jets in AGN, which is important if the AGN gamma-ray radiation is to be understood \citep[e.g.][]{barkovBBB,khangulyan2013}. 

Concerning other types of sources, the characterization of the radiation properties from jet-obstacle interactions could also shed light on the main factors shaping jet variability at high energies in microquasars \citep[e.g.][]{owocki2009}, or the post-shock emission in pulsar-wind termination shocks in some binary systems \citep[e.g.][]{bosch-ramon2013}. Regarding jets with strong magnetic fields with polarity changes, it has been proposed that the interaction with obstacles may lead to magnetic reconnection and efficient non-thermal emission \citep[see][]{bosch-ramon2012b}. This kind of interaction is very complex given the strong dynamical role of the magnetic field and its non-trivial topological properties and deserves a specific treatment, which is beyond the scope of this paper.

\section{The jet-RG wind interaction}\label{phys}

\subsection{Physical scenario}

The study presented here is based on the interaction of a jet with an RG surrounded by its dense and slow wind. The basic hydrodynamical ingredients of this study are: a highly relativistic, supersonic flow with parallel fluid lines on the scales of interest; a cold obstacle at rest; and a fraction of the internal energy downstream of the standing shock being in the form of non-thermal particles. The initial speed of the RG is taken to be $\ll c$ if it is to be approximated as at rest, so the problem can be considered axisymmetric around an axis crossing the star and with the jet flow direction. In addition, the RG wind must have enough inertia to stand the jet impact while it penetrates into the jet, which can be expressed as $\rho_{\rm rg}v_{\rm rg}^2>L_{\rm j}/\pi R_{\rm j}^2c$, 
where $\rho_{\rm rg}$ and $v_{\rm rg}$ are the RG wind density and velocity perpendicular to the jet, and $L_{\rm j}$ and $R_{\rm j}$ are the jet power and radius, respectively. In addition, in the standing shock scenario the velocity acquired by the RG wind due to jet impact must be $\ll c$. If these conditions are fulfilled, the problem becomes two dimensional, and the steady state of the physical system can be taken as the steady state of the shocked jet flow, which significantly lightens the hydrodynamical calculations presented in what follows.

\subsection{Simulations}

\begin{figure}
\includegraphics[width=0.45\textwidth,angle=0]{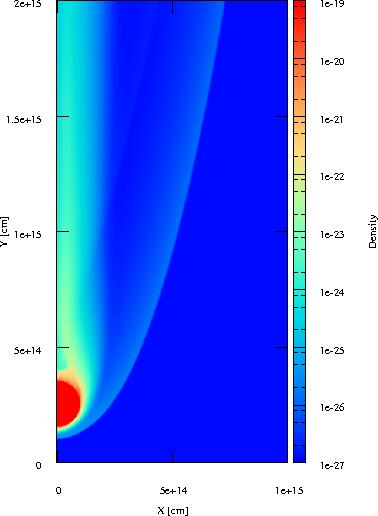}
\caption{Distribution of density in the flow frame in cgs units.}
\label{f0}
\end{figure}

The jet-RG wind interaction was simulated in 2D assuming axisymmetry. We solved the equations of relativistic hydrodynamics using a Marquina Riemann solver \citep{donat1996}, second-order spatial reconstruction scheme (minbee slope limiter, plus an additional dissipation term from \citealt{col85}), and an adiabatic gas with index $\hat\gamma=4/3$. The resolution is 600 cells in the vertical direction, the $z$-axis, and 300 in the radial direction, the $r$-axis, and the physical size is $z_{\rm max}=2\times 10^{15}$~cm in the $z$-direction, and $r_{\rm max}=10^{15}$~cm in the $r$-direction.  The jet power injected in the grid is $\approx 10^{37}$~erg~s$^{-1}$, with a Lorentz factor $\gamma_{\rm j}=10$. We fix the RG wind radius to $R_{\rm rg}=10^{14}$~cm. A snapshot of the density distribution in the steady state is presented in Fig.~\ref{f0}. The shocked jet material is clearly seen in light blue, forming a cometary tail pointing upwards and surrounding the RG wind, red with green borders.

For simplicity, we modelled the RG wind region as a sphere with a density $10^8$ times that of the jet in the FF. As the shocked RG wind is much slower than the shocked jet flow, we considered it static, and thus the steady state of the simulation is quickly reached. In a more realistic setup, there would be more wind momentum flux perpendicular to the jet direction, slightly widening the shocked flow structure, but this setup would have required a simulation $\sim 10^3$ times longer.

We chose the grid resolution such that numerical viscosity played a negligible role in the generation of internal energy in the shocked flow. There is some mixing in the RG wind-shocked jet flow boundary, but it has a minor impact on the radiation estimates presented below. This mixing slows down the shocked jet flow close to the axis, reducing the effect of Doppler boosting and thus weakening the emission from those (small) regions. 

For illustrative purposes, the simulation result is interpreted as an AGN jet of $L_{\rm j}\approx 10^{44}$~erg~s$^{-1}$ and $R_{\rm j}=1$~pc, much larger than the grid, interacting with an RG wind region whose size is derived assuming jet-stellar wind pressure balance
\begin{equation}
R_{\rm rg}\approx \sqrt{\dot{M}_{\rm w}v_{\rm w}c/4L_{\rm j}}\times R_{\rm j}\approx 10^{14}\,{\rm cm},
\label{ro}
\end{equation}
for a stellar mass loss $\dot{M}_{\rm w}=10^{-8}\,M_\odot$/yr and a wind velocity $v_{\rm w}\approx 2\times 10^7$~cm~s$^{-1}$, typical of an RG. Note that for a real RG wind the actual size of the obstacle would be slightly larger than for the case of an obstacle of very large inertia, as the shocked wind region should be added in Eq.~(\ref{ro}).

Relativistic jets usually present half semi-opening angles of several degrees, $\alpha_{\rm j}\sim 0.1$~rad, which for a supersonic jet without lateral causal contact yields $\gamma_{\rm j}=1/\alpha_{\rm j}\sim 10$ \citep[see][and references therein]{khangulyan2013}. This condition implies an interaction height of $z_{\rm j}=\gamma_{\rm j}R_{\rm j}\sim 10$~pc.

\section{Non-thermal emission}

We use the hydrodynamical information obtained in Sect.~\ref{phys} to derive some important physical properties that allow for the characterization of the non-thermal emission. These properties are the radiative efficiency, the distribution of the non-thermal energy, the Doppler boosting, and the effective section of the RG wind as an obstacle for the jet.

\subsection{Radiative efficiency and non-thermal energy distribution}

To estimate the non-thermal emission, the radiative efficiency of the shocked flow is needed. This efficiency can be quantified as the radiative ($1/t_{\rm rad}$) to total ($1/t_{\rm c}$) cooling rate ratio ($f_{\rm rad}$):
\begin{equation}
f_{\rm rad}=t_{\rm rad}^{-1}/t_{\rm c}^{-1}=t_{\rm rad}^{-1}/(t_{\rm rad}^{-1}+t_{\rm nrad}^{-1})=(1+t_{\rm rad}/t_{\rm nrad})^{-1}\,,
\end{equation}
where $1/t_{\rm c}$ also takes non-radiative cooling, $1/t_{\rm nrad}$, into account. The timescale for non-radiative cooling, i.e. particle escape and adiabatic losses, is estimated in the flow frame (FF) as the distance between each point to the RG location ($d_{\rm rg}$) divided by the local velocity ($v_{\rm sh}$) times the shocked flow Lorentz factor ($\gamma_{\rm sh}$):
$t_{\rm nrad}=d_{\rm rg}/v_{\rm sh}\gamma_{\rm sh}$. 
For the radiative timescale we adopt: $t_{\rm rad}=C_{\rm IC}\,u^{-1}E^{-1}$,
where $C_{\rm IC}\approx 26$ in cgs units. This timescale is derived assuming IC in Thomson, and taking a reference electron energy of $E_{\rm IC}=m_e^2c^4/\epsilon_0$, where $E_{\rm IC}$ is roughly the energy of the IC cross-section maximum, and $\epsilon_0$ is the target photon energy and $u$ the energy density of a photon field coming from the RG, all in the FF. An effective stellar temperature of 3000~K has been taken, typical for an RG. The star luminosity, needed to obtain $u$, has been determined through $L_*=\dot{M}_{\rm w}v_{\rm w}c$ (LF). This is a rather conservative assumption, as in fact only a fraction of all the star radiation momentum goes to the stellar wind. Note that protons are not taken into account as emitting particles.

\begin{figure}
\includegraphics[width=0.45\textwidth,angle=0]{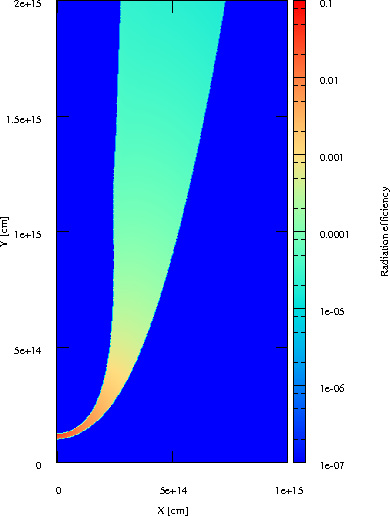}
\caption{Distribution of radiation efficiency for the shocked jet region.}
\label{f1}
\end{figure}

\begin{figure}
\includegraphics[width=0.45\textwidth,angle=0]{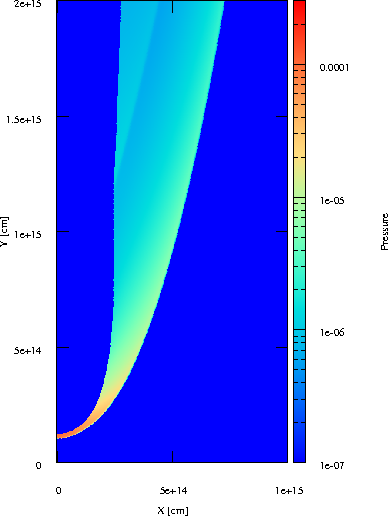}
\caption{Distribution of pressure in cgs units for the shocked jet region.}
\label{f2}
\end{figure}

Figure~\ref{f1} shows the distribution of $f_{\rm rad}$ in the shocked flow. As seen in the figure, for the adopted jet and RG wind properties, $f_{\rm rad}$ is everywhere below one for the case simulated, i.e. the flow is adiabatic. In this case, the thermal pressure ($P$) of the shocked flow is a good proxy for the non-thermal energy distribution, which can be taken to be proportional to $P$ all through the shocked flow. In Fig.~\ref{f2}, the distribution of $P$ in the shocked flow is shown in a colour scale, from red to light green.

The characterization of the radiative and the non-radiative timescales, and of the non-thermal energy distribution, allows for the computation of an approximate total luminosity for IC in the FF, $L_{\rm IC}$, simplifying the IC emissivity as that at $E_{\rm IC}$. For moderately magnetized shocked flows, the synchrotron luminosity $L_{\rm sync}$ may be higher than $L_{\rm IC}$, as $L_{\rm sync}\propto E_{\rm max}^{3-\alpha}$, where $\alpha$ is the power-law index of the particle energy distribution, and $E_{\rm max}$ is the maximum electron energy, which could be $\gg E_{\rm IC}$. Under efficient particle acceleration and $\gamma_{\rm j}\gtrsim 10$, the synchrotron spectrum may contribute to or dominate in the gamma-ray band. However, as the magnetic field is not well constrained, we will assume here that the synchrotron component can be neglected. 

The adopted radiation cooling model is simple, although it accounts for the basic IC cross-section properties, and the target density behaviour under relativistic motion of the emitter. Therefore, the model should yield reasonable luminosity predictions for IC gamma rays, if gamma-ray pair-creation absorption is negligible. 

\subsection{Doppler boosting and obstacle effective radius}

The radiation enhancement due to Doppler boosting is 
\begin{equation}
\delta_{\rm sh}^4=1/\gamma_{\rm sh}^4(1-\cos \theta_{\rm sh}\beta_{\rm sh})^4,
\end{equation}
which can be approximated as $\approx 16\gamma_{\rm sh}^4$ when the angle between the jet and the observer directions is $\theta_{\rm obs}\ll 1/\gamma_{\rm sh}$.

\begin{figure}
\includegraphics[width=0.42\textwidth,angle=0]{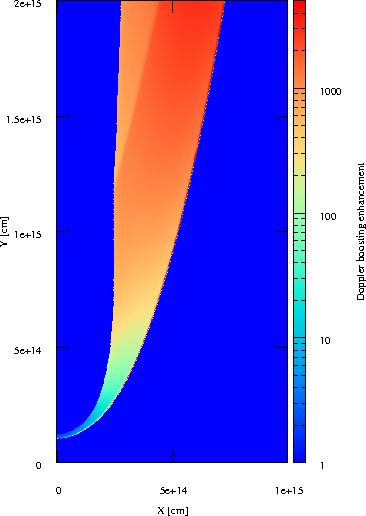}
\caption{Distribution of the Doppler boosting enhancement ($\delta_{\rm sh}^4$) for the shocked jet region.}
\label{f3}
\end{figure}

Figure~\ref{f3} shows the distribution of the factor of enhancement of the emission from the shocked flow because of Doppler boosting. As seen in Fig.~\ref{f3}, values of $\delta_{\rm sh}^4$ of a few are found right after the shock, but further downstream $\delta_{\rm sh}^4$ achieves values as high as several thousand.

\begin{figure}
\includegraphics[width=0.4\textwidth,angle=0]{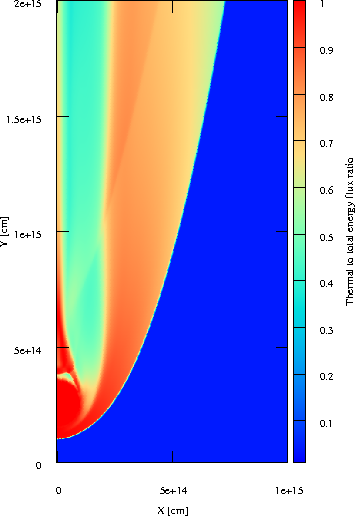}
\caption{Distribution of the ratio of fluxes of thermal energy to kinetic plus thermal energy.}
\label{f4}
\end{figure}

Another important result from our simulation is derived when computing $\dot{U}_{\rm sh}$, the rate of the total thermal energy generated in the interaction region in the LF, where the total thermal energy is the sum of internal energy plus the pressure component. The value of $\dot{U}_{\rm sh}$ has been computed as the fraction of $L_{\rm j}$ that escapes the grid in the form of thermal energy. From this, we can derive the effective radius ($R_{\rm eff}$) of the RG wind when acting as an obstacle for the jet
\begin{equation}
\dot{U}_{\rm sh}=(R_{\rm eff}/R_{\rm j})^2L_{\rm j}\approx 4\times 10^{36}\,{\rm erg~s}^{-1},
\end{equation}
which turns out to be $R_{\rm eff}\approx 6\,R_{\rm rg}$ because of effective kinetic to internal energy conversion even in regions far from the RG. This effective conversion manifests itself locally as well, as shown by Fig.~\ref{f4}. 

The value of $R_{\rm eff}$ is actually dependent on $R_{\rm rg}$ and $z_{\rm max}$. To check this dependence, two additional simulations were carried out, both with twice the cell size of the presented simulation, one with the same grid size, and the other with a grid that is twice as large, and both with the same remaining parameters. These simulations show that $R_{\rm eff}$ grows slowly with $r_{\rm max}$. This is expected to happen as the oblique shock becomes weaker farther away from the RG, but the actual weakening of the shock is sensitive to the specific value of the unshocked flow internal energy, which is unknown. Note also that to derive $R_{\rm eff}$ all the grid is taken into account; in reality, the efficiency of energy conversion is larger if the grid outermost region without shock presence is neglected. Therefore, $R_{\rm eff}=6\,R_{\rm rg}$ may be considered a lower limit. 

\section{The IC luminosity from a jet-obstacle interaction}\label{res}

\subsection{The IC luminosity to the observer}

Using the hydrodynamics results and under the simplified emission model described, we can derive an approximate observer IC luminosity of the shocked flow within the grid from
\begin{equation}
L_{\rm IC,obs}\sim\int_{\rm sh}\delta_{\rm sh}^4(1-\cos{\theta_{\rm *obs}})\eta_{\rm NT}\hat\gamma P(\hat\gamma-1)^{-1}t_{\rm rad}^{-1}dv_{\rm sh},
\label{icobs}
\end{equation}
where $dv_{\rm sh}$ is the shocked flow differential volume, and $\theta_{\rm *obs}$ the angle between the stellar photons and the observer direction, both in the FF. The term $(1-\cos{\theta_{\rm *obs}})$ is a term related to the IC scattering probability-dependence on the stellar photon and observer directions. The quantity $\eta_{\rm NT}$ is the ratio of non-thermal to total thermal energy density; it is a free parameter of value between 0 and 1. For a typical accelerated electron population, say with a particle energy distribution harder than $\propto E^{-3}$, given the value of $\epsilon_0$, and accounting for Doppler boosting, the IC radiation will be mostly radiated in the GeV--TeV range, as seen by the observer. 

From Eq.~(\ref{icobs}), the total observer IC luminosity is computed, obtaining $L_{\rm IC,obs}\sim 10^{35}\eta_{\rm NT}\,{\rm erg~s}^{-1}$,
which is $\approx 1\,\eta_{\rm NT}$\% of the jet luminosity injected in the grid, and a $\approx 3\,\eta_{\rm NT}$\% of $\dot{U}_{\rm sh}$. On the other hand, neglecting relativistic effects and confining the emitter to the RG wind vicinity, one would get $L_{\rm IC,obs}\sim 3\times 10^{32}\eta_{\rm NT}\,{\rm erg~s}^{-1}$,
which is $\sim 300$ times smaller than the more realistic value. This shows that the radiation can be underestimated if the interaction region is not globally taken into account. 

\begin{figure}
\includegraphics[width=0.4\textwidth,angle=0]{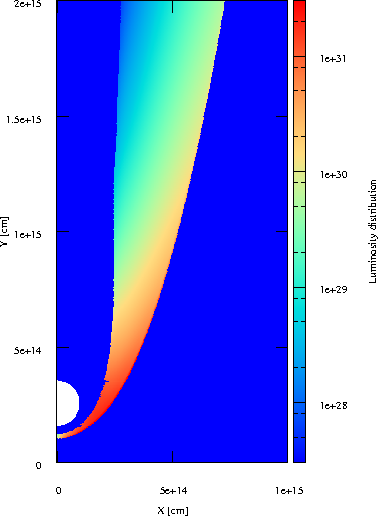}
\caption{Distribution of the observer IC luminosity in cgs units for the shocked jet region.}
\label{f5}
\end{figure}

Figure~\ref{f5} shows the structure of the observer IC emitter for $\theta_{\rm obs}\ll 1/\gamma_{\rm sh}$, produced in ring-like cells of volume $2\pi r_{\rm j}\Delta r_{\rm j}\Delta z_{\rm j}$, where $\Delta r_{\rm j}=r_{\rm max}/300$ and $\Delta z_{\rm j}=z_{\rm max}/600$. As seen in the figure, most of the emission comes from the closest regions to the RG, but also from farther away, close to the jet shock.

\subsection{Generalization of the results}

For the adopted approximation $z_{\rm j}\sim \gamma_{\rm j}R_{\rm j}$, our results correspond to an emitter located at relatively high heights ($\sim 10$~pc) in a jet of moderate power ($\sim 10^{44}$~erg~s$^{-1}$). These results can nevertheless be generalized through
\begin{equation}
L_{\rm IC}\propto t_{\rm nrad}/t_{\rm rad}\sim L_{\rm j}z_{\rm j}^{-1}\,,
\label{coolz}
\end{equation}
so $L_{\rm IC,obs}$ will grow closer to the jet base and/or with a more powerful jet.
In saturation,  Eq.~(\ref{coolz}) is not valid anymore as $t_{\rm rad}\rightarrow t_{\rm nrad}$, and thus Eq.~(\ref{icobs}) becomes
\begin{equation}
L_{\rm IC,obs}^{\rm max}\sim 10^2\eta_{\rm NT}\dot{U}_{\rm sh}\approx 4\times 10^{38}\eta_{\rm NT}\,{\rm erg~s}^{-1}, 
\end{equation}
i.e. it gets constant, for the same remaining parameters: $\gamma_{\rm j}$ and $\dot{M}v_{\rm w}$.

The value of $L_{\rm IC,obs}$ is also $\propto \gamma_{\rm j}^2$, for $\theta_{\rm obs}\ll 1/\gamma_{\rm sh}$ and $\gamma_{\rm j,sh}\gg 1$. This relation comes from the relation $\gamma_{\rm sh}\propto \gamma_{\rm j}$, strictly valid in the highly relativistic case, and the link between the IC luminosity in the LF ($L_{\rm IC,lab}$) and $L_{\rm IC,obs}$:
\begin{equation}
L_{\rm IC,obs}/L_{\rm IC,lab}\propto\ \delta_{\rm sh}^4/\gamma_{\rm sh}^2\approx 16\,\gamma_{\rm sh}^2\propto \gamma_{\rm j}^2.
\label{dopsh}
\end{equation}
Therefore, the computed estimates can be analytically generalized through 
\begin{equation}
L_{\rm IC,obs}\propto \gamma_{\rm j}^2R_{\rm rg}^2L_{\rm j}z_{\rm j}^{-1}\propto \gamma_{\rm j}^2\dot{M}v_{\rm w}L_{\rm j}z_{\rm j}^{-1},
\end{equation}
or in saturation, through
\begin{equation}
L_{\rm IC,obs}\propto \gamma_{\rm j}^2\dot{M}v_{\rm w},
\end{equation}
relations that also account for different types of winds via $\dot{M}v_{\rm w}$. 

For a microquasar jet, given the typical clump sizes \citep[e.g.][]{owocki2009}, clumps with $R_{\rm eff}\sim R_{\rm j}$ may reach the jet. Also, IC emission may take place in saturation given the proximity of the stellar companion, yielding luminosities as high as
\begin{equation}
L_{\rm IC,obs}\sim 16\gamma_{\rm sh}^2\eta_{\rm NT}L_{\rm j}, 
\end{equation}
which could be $\gg L_{\rm j}$ for some relatively exceptional, and brief, events.

\section{A population of red giants within an AGN jet}

We explore in what follows whether collectively RG may be relevant for the non-thermal emission from AGN jets.
About a 5\% of the stars in the AGN vicinity are expected to be RG \citep{syer1999}. In addition, there are observational hints of stars interacting with AGN jets \citep[e.g.][]{har2003,muller2014}. Therefore, the scenario explored in this work, and already postulated as the origin of high-energy emission by several authors (see Sect.~\ref{intro}), could be actually a rather common phenomenon. 

We take the nearby, gamma-ray emitting AGN \object{M87} as an example. With $\sim 10^{11}$ stars in the inner $\sim 1$~kpc of the galaxy \citep[see fig.~7 in][]{gebhardt2009}, the number of RG within one jet should be about
\begin{equation}
N_{\rm RG}\sim 0.05\times 10^{11}\times 3\alpha_{\rm j}^2/4\approx 4\times 10^7\,,
\end{equation}
where we take $\alpha_{\rm j}\sim 0.1$, and $3\alpha_{\rm j}^2/4$ is the jet to total volume ratio. 

Assuming an RG density $n_{\rm RG}\propto 1/z_{\rm j}^\xi$, an average $\dot{M}_{\rm w}v_{\rm w}\sim 10^{25}$~g~cm~s$^{-2}$, the relation from Eq.~(\ref{coolz}), negligible synchrotron radiation, a constant $\gamma_{\rm j}=10$, and $L_{\rm j}=10^{44}$~erg~s$^{-1}$, each interaction would produce in gamma rays
\begin{equation}
L_{\rm IC,obs}(z_{\rm j})\sim L_{\rm IC,obs}(z_{\rm min})\times (z_{\rm min}/z_{\rm j})\,,
\end{equation}
where $z_{\rm min}\approx 8\times 10^{15}$~cm is found through $L_{\rm IC,obs}({\rm 10~pc})=10^{35}\eta_{\rm NT}\,{\rm erg~s}^{-1}$, and fixing the $z_{\rm min}$-luminosity to the saturation value: $L_{\rm IC,obs}(z_{\rm min})=L_{\rm IC,obs}^{\rm max}$. All this allows us to derive
\begin{equation}
L_{\rm IC,obs}^{\rm tot}=\int_{\rm z_{\rm min}}^{\rm z_{\rm max}} n_{\rm RG}(z_{\rm j})L_{\rm IC,obs}(z_{\rm j})\,dv_{\rm j}=
\label{lmaxic}
\end{equation}
$$
=N_{\rm RG}\left(\frac{3-\xi}{2-\xi}\right)\left(\frac{z_{\rm max}^{2-\xi}-z_{\rm min}^{2-\xi}}{z_{\rm max}^{3-\xi}-z_{\rm min}^{3-\xi}}\right)L_{\rm IC,obs}(z_{\rm min})\times z_{\rm min}\,,
$$
where $dv_{\rm j}$ is the jet differential volume. 

The dominant contribution to $L_{\rm IC,obs}^{\rm tot}$ will come from the inner-jet regions, or from galactic scales, depending on whether $\xi$ is $>$ or $<$ than 2. For $\xi\approx 2$, the $z_{\rm min,max}$-dependence of $L_{\rm IC,obs}^{\rm tot}$ is weak, but the external jet regions dominate. For this $\xi$-value\footnote{A slope $\xi\sim 2$ can be derived from \cite{gebhardt2009} for kpc scales, becoming harder closer to the centre of \object{M87}.}, Eq.~(\ref{lmaxic}) gives $L_{\rm IC,obs}^{\rm max}\sim 5\times 10^{41}\eta_{\rm NT}$~erg~s$^{-1}$, implying that the contribution from all RG within 1~kpc could be up to a factor of a few above the persistent GeV--TeV luminosities in \object{M87} \citep[e.g.][]{abdo2009,magic2012}. It seems unlikely however that the acceleration efficiency will be $\eta_{\rm NT}\sim 1$, and we may be overestimating the average value of $\dot{M}v_{\rm w}$. On the other hand, we do not take into account the asymptotic giant branch (AGB) and massive star contribution, both of which should be added to that of RG. Non-stellar objects such as clouds could also be relevant. Finally, the value of $L_*$ could be significantly higher, at least in some stars. In conclusion, $L_{\rm IC,obs}^{\rm max}$ may explain the persistent GeV--TeV emission in \object{M87}, and could be significant for AGN jets in general. 

Note that the viewing angle with respect to the \object{M87} jet is not zero, but rather $\sim 20^\circ$ \citep[e.g.][]{acc09}. However, given the obtained values of $\gamma_{\rm sh}$ ($\lesssim 4$), an angle of $20^\circ$ is small enough for the condition $\theta_{\rm obs}\ll 1/\gamma_{\rm sh}$ to marginally apply. Here, the gamma-ray luminosity for \object{M87} has been computed using the proper viewing angle, although taking $\theta_{\rm obs}=0$ yields very similar results.

\section{Discussion}

The emission coming from the jet in Blazar AGN will be more boosted than that from the downstream region of a standing shock. Thus, strong jet activity could mask the putative jet-obstacle emission. However, for slightly misaligned, relatively nearby AGN, or those without significant non-thermal activity intrinsic of the jet, the contribution from populations of stars interacting with the jet could be important, $\sim 0.1-1$\% of $L_{\rm j}$ (but strongly dependent on $\eta_{\rm NT}$, $\gamma_{\rm j}$, and $\theta_{\rm obs}$). This collective emission should be constant unless $\xi$ were significantly softer than 2, which does not seem realistic. There could also be rare flaring events when a dense cloud, an AGB, a Wolf-Rayet or a luminous blue variable (WR/LBV) star, entered into the jet close to its base \citep[e.g.][]{araudo2010,barkov2010,hubbard2006,araudo2013}. In the case of microquasar jets, the ratio $R_{\rm eff}/R_{\rm j}$ can be larger than in the extragalactic case and fluctuations may be more common, with the arrival of larger clumps leading to brighter events \citep[][]{owocki2009,araudo2009}. This may also happen in binary systems hosting a non-accreting pulsar \citep[see][and references therein]{bosch-ramon2013,paredes-fortuny2015}.

The obtained numerical results have been used to characterize the effective section of the RG wind as an obstacle for the jet, the distribution of the non-thermal particles through the constant $\eta_{\rm NT}$, and the effect of Doppler boosting in the shocked flow for $\gamma_{\rm j}\gg 1$. Previous research focussed on a region downstream of the jet shock next to the obstacle and with a similar size \cite[e.g.][]{araudo2009,araudo2010,barkov2010,barkovAAA}, and Doppler boosting effects were not considered as the flow is mildly relativistic there. As shown here, both approximations can lead to the underestimation of the non-thermal emission from jet-obstacle interactions. 

Our work has some caveats. The simulation, in axisymmetry, without thermal cooling nor magnetic field, with modest resolution, and with a simple prescription of the RG wind, is not very realistic, although it captures the basic features of the shocked region of a hydrodynamic jet. The value of $f_{\rm rad}$ is a rough characterization of the radiative efficiency for IC, and our model of the IC emission is very simplified: it assumes IC only in Thomson, and accounts only for the electron energy at the maximum of the IC cross section. This simplification introduces an uncertainty in the predicted luminosity by a factor of a few. On the other hand, our radiation calculations account for the main features of IC scattering, and so we expect our results to be correct within an order of magnitude.
Note that the value $\eta_{\rm NT}$ cannot be derived from first principles, and despite $\eta_{\rm NT}\gtrsim 0.1$ are not uncommon in relativistic jet sources, $\eta_{\rm NT}\ll 1$ are also possible. It is also difficult to constrain the magnetic field strength in jets, which affects the calculation of synchrotron radiation. More accurate calculations will be presented in future work. 

\begin{acknowledgements} 
V.B-R. acknowledges support by the Spanish Ministerio de Econom\'{\i}a y Competitividad (MINECO) under grant AYA2013-47447-C3-1-P.
This research has been supported by the Marie Curie Career Integration Grant 321520.
V.B-R. also acknowledges financial support from MINECO and European Social Funds through a Ram\'on y Cajal fellowship.
\end{acknowledgements}

\bibliographystyle{aa}
\bibliography{text}
\end{document}